\documentclass[preprintnumbers,10pt,nofootinbib]{revtex4}

\usepackage{amsmath,latexsym,amssymb,amsfonts}
\usepackage[dvips]{graphicx}
\usepackage{bm}

\begin{document}


\title{\textbf{Dynamics of magnetic shells and information loss problem}}

\author{
\textsc{Bum-Hoon Lee$^{1,2}$}\footnote{{\tt bhl{}@{}sogang.ac.kr}},\;
\textsc{Wonwoo Lee$^{1}$}\footnote{{\tt  warrior{}@{}sogang.ac.kr}}
and \textsc{Dong-han Yeom$^{3}$}\footnote{{\tt innocent.yeom{}@{}gmail.com}}
}

\affiliation{
$^{1}$\small{Department of Physics, Center for Quantum Spacetime, Sogang University, Seoul 121-742, Republic of Korea}\\
$^{2}$\small{Asia Pacific Center for Theoretical Physics, Pohang 790-784, Republic of Korea}\\
$^{3}$\small{Leung Center for Cosmology and Particle Astrophysics, National Taiwan University, Taipei 10617, Taiwan}
}

\begin{abstract}
We investigate dynamics of magnetic thin-shells in three dimensional anti de Sitter background. Because of the magnetic field, an oscillatory solution is possible. This oscillating shell can tunnel to a collapsing shell or a bouncing shell, where both of tunnelings induce an event horizon and a singularity. In the entire path integral, via the oscillating solution, there is a non-zero probability to maintain a trivial causal structure without a singularity. Therefore, due to the path integral, the entire wave function can conserve information. Since an oscillating shell can tunnel after a number of oscillations, in the end, it will allow an infinite number of different branchings to classical histories. This system can be a good model of the effective loss of information, where information is conserved by a solution that is originated from gauge fields.
\end{abstract}

\maketitle

\newpage

\tableofcontents


\section{\label{sec:int}Introduction}

The information loss problem \cite{Hawking:1976ra} is one of the most interesting topics to understand the theory of everything as well as the principles of quantum gravity. The information loss problem is difficult to resolve since there appears a contradiction between well-experienced principles. These contradictory principles are basically \textit{unitary quantum mechanics} and \textit{general relativity}. There was a trial to resolve the tension between unitary quantum mechanics and general relativity, the black hole complementarity principle \cite{Susskind:1993if}, though recently it was widely accepted that this principle is inconsistent \cite{Yeom:2008qw,Almheiri:2012rt} and hence we have to give up one of the principles, e.g., either unitarity or general relativity.

\paragraph{Effective loss of information} The theory of quantum gravity should explain the tension between unitarity and general relativity. Although there are various candidates to try to resolve this trouble, one of the promising candidates is the \textit{effective loss of information} \cite{Hawking:2005kf,Sasaki:2014spa}.

If we interpret this idea in the Lorentzian signatures \cite{Sasaki:2014spa} (also, recently \cite{Hartle:2015bna}), there exists a \textit{unitary observer} who gathers all information in the \textit{superspace} (a functional space of fields). This means that the unitary observer should see approximately a superposition of various semi-classical histories, including a non-trivial geometry (so to speak $g^{(1)}_{\mu\nu}$, that describes an evaporating black hole according to Hawking's calculations) and a trivial geometry (so to speak $g^{(2)}_{\mu\nu}$, hence no event horizon and no singularity). Each history should be weighted by a proper probability:
\begin{eqnarray}
\langle g_{\mu\nu} \rangle \simeq p_{1} g^{(1)}_{\mu\nu} + p_{2} g^{(2)}_{\mu\nu} + ... .
\end{eqnarray}
If there exists a trivial geometry among semi-classical histories, then correlations between the collapsing matter and the asymptotic infinity will be conserved following this history: $\langle \phi \phi \rangle_{2} \rightarrow \mathrm{const}$ as time goes on. On the other hand, correlations will decay to zero for the case of the non-trivial geometry that has an event horizon and a singularity: $\langle \phi \phi \rangle_{1} \rightarrow 0$ as time goes on. Then the entire correlation function will be dominated by the trivial geometry in the end although its probability in itself is more highly suppressed than the most probable and non-trivial geometry, i.e., $p_{1} \gg p_{2}$:
\begin{eqnarray}
\langle \phi \phi \rangle \simeq p_{1} \langle \phi \phi \rangle_{1}+ p_{2} \langle \phi \phi \rangle_{2} + ...  \simeq_{t \rightarrow \infty} p_{2} \times \mathrm{const}.
\end{eqnarray}
This explains the information conservation of the unitary observer \cite{Maldacena:2001kr}. In the end, the unitary observer will see a superposition of trivial geometry and non-trivial geometry. Although each history (e.g., $g^{(1)}_{\mu\nu}$ or $g^{(2)}_{\mu\nu}$) will satisfy the semi-classical equations, the superposed geometry (i.e., $\langle g_{\mu\nu} \rangle$) does not necessarily satisfy the semi-classical equations. Therefore, for the unitary observer, he/she does not necessarily satisfy general relativity. This violation can be even large around the horizon scale, and hence one may see some strange behaviors around the horizon scale, where this can be named by a firewall phenomenon and the effects of the firewall should be observed by the unitary observer as expected by \cite{Hwang:2012nn}; of course, it does not necessarily mean that this is a kind of barrier where a free-falling observer cannot penetrate there.

On the other hand, a semi-classical observer (who lives in a specific metric) should follow one of histories, since each history is classicalized independently. Probably, the semi-classical observer will see the most probable history $g^{(1)}_{\mu\nu}$, where it will contain an evaporating black hole geometry according to well-known semi-classical quantum field theory. Therefore, the semi-classical observer will see general relativity and the local quantum field theory, while also experience a violation of unitarity. This is the reason why we call this idea the effective loss of information; information is conserved, but semi-classical observers probably cannot restore information effectively.

\paragraph{Motivations} This line of understanding elegantly explains the tension between unitary quantum mechanics and general relativity. However, this requires technical justifications. The critical question is as follows: \textit{for all gravitational collapses or formations of black objects, can there exist a trivial and classicalized geometry via non-perturbative effects?} Regarding this question, we do not have a general proof on the existence of such non-perturbative channels for \textit{all} possible gravitational collapses. However, what we can say is that, at least, for some \textit{special} cases, we can explicitly construct such non-trivial and trivial geometries. One of special examples is a thin-shell collapsing bubble \cite{Israel:1966rt} in anti de Sitter (AdS) background \cite{Sasaki:2014spa}. Even though we are still considering a special example, as we investigate further, we believe that we can extend to the most general case, as long as our understanding is fundamentally true.

For the thin-shell collapsing case, it clearly shows that there can exist a tunneling channel that mediates a trivial geometry. This is nothing but a tunneling from a collapsing shell to a bouncing shell. Therefore, in the end, the tunneling channels are not so complicated. However, can there be any other complication of such a tunneling? If the shell dynamics allows more tunneling channels, we may need to sum over much more divergent histories. To find a good toy model, in this paper, we consider a thin-shell (in fact, thin-ring in $2+1$ dimensions) that has a magnetic field on the shell, motivated from previous works in \cite{Lee:2013ega}. Because of fluxes, there is a competition between various factors: vacuum energy pressure, gravitational attraction, and conservation of magnetic fluxes. This will allow more complicated and rich histories, and hence this model will be an interesting toy model that realizes the effective loss of information.

\paragraph{Contents} This paper is organized as follows. In SEC.~\ref{sec:mod}, we discuss a thin magnetic shell system in three dimensions. In SEC.~\ref{sec:dyn}, we summarize possible shell dynamics as well as their tunneling processes. Finally, in SEC.~\ref{sec:con}, we summarize our results and discuss possible future applications. In this paper, we use the conventions: $c = \hbar = G = 1$.

\section{\label{sec:mod}Three-dimensional magnetic shells}

To realize a thin magnetic shell, we introduce a three dimensional system with a complex scalar field and a gauge field. Since the system is in three dimensions, there can exist a field combination that has a constant magnetic field within a certain region. This is partly similar with a vortex \cite{Lee:2013ega}, though we will consider the case when the magnetic field is confined on the shell, where this is different from the case of a vortex.

\subsection{Model}

Let us introduce the action in three dimensions by
\begin{eqnarray}
S = \int d^{3}x \sqrt{-g} \left[ \frac{1}{16 \pi} \left( R - 2\Lambda \right) - \frac{1}{2}\left(\phi_{;\mu}-ieA_{\mu}\phi \right)g^{\mu\nu}\left(\overline{\phi}_{;\nu}+ieA_{\nu}\overline{\phi}\right) - U(\phi\bar{\phi}) -\frac{1}{16\pi}F_{\mu\nu}F^{\mu\nu} \right],
\end{eqnarray}
where $R$ is the Ricci scalar, $\phi$ is a complex scalar field with a gauge coupling $e$ and a gauge field $A_{\mu}$, where $F_{\mu\nu}=A_{\nu;\mu}-A_{\mu;\nu}$.

The Einstein equation is
\begin{eqnarray}
G_{\mu\nu} + \Lambda g_{\mu\nu} = 8\pi T_{\mu\nu},
\end{eqnarray}
where the energy-momentum tensor is
\begin{eqnarray}
T_{\mu\nu} &=& \frac{1}{2}\left(\phi_{;\mu}\overline{\phi}_{;\nu}+\overline{\phi}_{;\mu}\phi_{;\nu}\right) +\frac{1}{2}\left(\phi_{;\mu}ieA_{\nu}\overline{\phi}-\overline{\phi}_{;\nu}ieA_{\mu}\phi-\overline{\phi}_{;\mu}ieA_{\nu}\phi+\phi_{;\nu}ieA_{\mu}\overline{\phi}\right)
\nonumber \\
&& {}+\frac{1}{4\pi}F_{\mu \rho}{F_{\nu}}^{\rho}+e^{2}A_{\mu}A_{\nu}\phi\overline{\phi}+\mathcal{L}g_{\mu \nu},
\end{eqnarray}
where $\mathcal{L} = - (1/2)\left(\phi_{;\mu}-ieA_{\mu}\phi \right)g^{\mu\nu}\left(\overline{\phi}_{;\nu}+ieA_{\nu}\overline{\phi}\right)-U -(1/16\pi)F_{\mu\nu}F^{\mu\nu}$.

The field equations for the complex scalar field $\phi$ and the gauge field $A_{\mu}$ are as follows:
\begin{eqnarray}
\label{eq:phi}\phi_{;\mu\nu}g^{\mu\nu}-ieA^{\mu}\left(2\phi_{;\mu}-ieA_{\mu}\phi\right)-ieA_{\mu;\nu}g^{\mu\nu}\phi -\frac{dU}{d\bar{\phi}} &=& 0,
\\
\label{eq:A}\frac{1}{2\pi}{F^{\nu}}_{\mu;\nu}+ie\phi\left(\overline{\phi}_{;\mu}+ieA_{\mu}\overline{\phi}\right)-ie\overline{\phi}\left(\phi_{;\mu}-ieA_{\mu}\phi\right) &=& 0.
\end{eqnarray}

\subsection{Static and circular symmetric metric}

We use the static and circular symmetric metric ansatz
\begin{eqnarray}
ds^{2} = - h(r) dt^{2} + \frac{1}{f(r)} dr^{2} + r^{2} d\varphi^{2}.
\end{eqnarray}
In addition, we impose
\begin{eqnarray}
\phi(r,\varphi) = \frac{\psi(r)}{\sqrt{4\pi}} e^{in\varphi},
\end{eqnarray}
where $n$ is an integer and $\psi(r)$ is a real valued function. Since we assume that $\psi$ is real and there is the angular dependence, the gauge field is sufficient to choose
\begin{eqnarray}
A_{\mu} = \left[0,0,\frac{n(1-a)}{e}\right].
\end{eqnarray}

We can reduce the components of the action as follows:
\begin{gather}
\phi_{;\mu}-ieA_{\mu}\phi = \left(
  \begin{array}{c}
    0\\
    \psi'\\
    in\psi a\\
  \end{array}
\right) \frac{e^{in \varphi}}{\sqrt{4\pi}},
\end{gather}
\begin{eqnarray}
\frac{1}{2} \left(\phi_{;\mu}-ieA_{\mu}\phi \right)g^{\mu\nu}\left(\overline{\phi}_{;\nu}+ieA_{\nu}\overline{\phi}\right) = \frac{1}{8\pi}\psi'^{2} f + \frac{n^{2} \psi^{2} a^{2}}{8\pi r^{2}},
\end{eqnarray}
\begin{gather}
F_{\mu\nu} = \frac{n a'}{e} \left(
  \begin{array}{ccc}
    0 & 0 & 0\\
    0 & 0 & -1 \\
    0 & 1 & 0 \\
  \end{array}
\right),
\end{gather}
and
\begin{eqnarray}
\frac{1}{16\pi} F_{\mu\nu}F^{\mu\nu} = \frac{n^{2} a'^{2}f}{8\pi e^{2} r^{2}},
\end{eqnarray}
where $'$ is a derivation with respect to $r$. Then, the reduced action becomes
\begin{eqnarray}
S &=& - \frac{1}{8} \int drdt \left[ \left( \frac{2rh''hf - rh'^2 f + rh'f'h + 2fh'h + 2f' h^{2}}{2h^{2}} + 2 \Lambda r \right) \right.\nonumber \\
&& \left. + 2 r\left( \psi'^{2} f + \frac{n^{2} a'^{2} f}{e^{2}r^{2}} + 8\pi U(\psi^{2}) + \frac{n^{2}\psi^{2} a^{2}}{r^{2}} \right) \right].
\end{eqnarray}

The components of Einstein tensors and energy-momentum tensors are
\begin{eqnarray}
G_{tt} &=& - \frac{h f'}{2r},\\
G_{rr} &=& \frac{h'}{2rh},\\
G_{\varphi \varphi} &=& \frac{2h'' hf - h'^{2}f + h' f' h}{4h^{2}}r^{2},\\
8\pi T_{tt} - g_{tt} \Lambda &=& \left( \psi'^{2} f + \frac{n^{2} a'^{2} f}{e^{2} r^{2}} + \frac{n^{2} \psi^{2} a^{2}}{r^{2}} + \Lambda + 8\pi U \right) h,\\
8\pi T_{rr} - g_{rr} \Lambda &=& \left( \psi'^{2} f + \frac{n^{2} a'^{2} f}{e^{2} r^{2}} - \frac{n^{2} \psi^{2} a^{2}}{r^{2}} - \Lambda - 8\pi U \right) \frac{1}{f},\\
8\pi T_{\varphi \varphi} - g_{\varphi \varphi} \Lambda &=& \left(- \psi'^{2} f + \frac{n^{2} a'^{2} f}{e^{2}r^{2}} + \frac{n^{2}\psi^{2}a^{2}}{r^{2}} - \Lambda - 8\pi U  \right)r^{2}.
\end{eqnarray}
Then one can completely write the Einstein equations: $G_{\mu\nu} + \Lambda g_{\mu\nu} = 8\pi T_{\mu\nu}$.
The other field equations are as follows:
\begin{eqnarray}
\psi'' + \left( \frac{f'}{f} + \frac{1}{r} \right) \psi' - \frac{n^{2} a^{2}}{r^{2}f} \psi - \frac{4\pi U'}{f} &=& 0, \label{eq:01}\\
a'' + \left( \frac{f'}{f} - \frac{1}{r} \right) a' - \frac{e^{2} \psi^{2}}{f} a &=& 0. \label{eq:02}
\end{eqnarray}

By comparing the $tt$-component and $rr$-component of the Einstein equations, we obtain one consistency relation:
\begin{eqnarray}
\psi'^{2} + \frac{n^{2} a'^{2}}{e^{2} r^{2}} = \frac{1}{4r} \left( \frac{h'}{h} - \frac{f'}{f} \right).
\end{eqnarray}
Therefore, if $f = h$, then $\psi' = a' = 0$ is required and hence all non-trivial solutions should be time-dependent.


\subsection{Thin-shell approximation}

To consider a non-trivial field combination, either we should consider fully dynamical field combinations by a numerical method or we need to use the thin-shell approximation \cite{Mann:2006yu} (although it is better to say thin-ring) so that the outside the shell and the inside the shell satisfy $\psi' = a' = 0$ (with different mass and vacuum energy) while $\psi'$ and $a'$ are non-trivial only on the shell.

For a static metric, if $\psi' = a' =0$, then $f=h$ and hence the following is a reasonable ansatz:
\begin{eqnarray}
ds^{2} = - f_{\pm}(R) dT^{2} + \frac{1}{f_{\pm}(R)} dR^{2} + R^{2} d\varphi^{2},
\end{eqnarray}
where the shell locates at $R = r(t)$, $-$ is for inside ($r > R$), and $+$ is for outside ($r < R$) the shell. The shell metric is represented by the reduced metric
\begin{eqnarray}
ds_{\mathrm{shell}}^{2} = - dt^{2} + r^{2}(t) d\varphi^{2}.
\end{eqnarray}

For inside and outside the shell, we choose the stationary solution:
\begin{eqnarray}
\psi_{-}(R) = 0, \;\;\; a_{-}(R) = 1, \;\;\; U'(\psi_{-}) = 0, \;\;\; \Lambda + 8\pi U(\psi_{-}) \equiv \Lambda_{-},
\end{eqnarray}
and
\begin{eqnarray}
\psi_{+}(R) = \psi_{\mathrm{f}}, \;\;\; a_{+}(R) = 0, \;\;\; U'(\psi_{+}) = 0, \;\;\; \Lambda + 8\pi U(\psi_{+}) \equiv \Lambda_{+}.
\end{eqnarray}
Then, for inside and outside, the metric should satisfy the equations:
\begin{eqnarray}
f_{\pm}'' + 2 \Lambda_{\pm} &=& 0,\\
f_{\pm}' + 2 r \Lambda_{\pm} &=& 0.
\end{eqnarray}
Therefore,
\begin{eqnarray}
f_{\pm} = - M_{\pm} + \frac{R^{2}}{\ell_{\pm}^{2}},
\end{eqnarray}
where $\ell_{\pm} = 1/\sqrt{-\Lambda_{\pm}}$ and this is well-embedded in one type of Banados-Teitelboim-Zanelli (BTZ) solutions \cite{Banados:1992wn}. In this paper, we choose $M_{-} = -1$ (pure AdS) and $M_{+} > 0$ to maintain a trivial geometry for inside the shell.

The junction equation and the energy-conservation equation become \cite{Mann:2006yu}
\begin{eqnarray}
\epsilon_{-}\sqrt{\dot{r}^{2} + f_{-}} - \epsilon_{+}\sqrt{\dot{r}^{2} + f_{+}} &=& \lambda r,\\
\frac{d}{dr} \left( r \lambda \right) + p &=& 0,
\end{eqnarray}
where $\lambda$ is the tension, $p$ is the pressure of the shell, $\epsilon_{\pm} = \pm 1$, and the signs of $\epsilon_{\pm}$ should be recovered by comparing with the signs of extrinsic curvatures. Especially, the first equation is simplified to
\begin{eqnarray}
\dot{r}^{2} + V(r) &=& 0,\\
V(r) &\equiv& f_{+} - \frac{\left( f_{-} - f_{+} - \lambda^{2} r^{2} \right)^{2}}{4\lambda^{2} r^{2}}.
\end{eqnarray}

\subsection{Details on tensions}

One important point from the magnetic shell is that the relation between the tension and the pressure can be functions of $r$. The equation of state should contain information or back-reactions on $n$ and $e$, etc. The kinetic terms in the thin-shell limit ($\Delta r \ll r$) becomes
\begin{eqnarray}
\psi'^{2} \simeq \left( \frac{\Delta \psi}{\Delta r}  \right)^{2} \simeq \frac{\psi_{\mathrm{f}}^{2}}{\Delta r^{2}} &\propto& \delta(R-r),\\
\frac{1}{e^{2}} a'^{2} \simeq \frac{1}{e^{2}} \left( \frac{\Delta a}{\Delta r}  \right)^{2} \simeq \frac{1}{e^{2}} \frac{1}{\Delta r^{2}} &\propto& \delta(R-r),
\end{eqnarray}
by assuming $\psi_{\mathrm{f}}^{2}/\Delta r$ and $1/(e^{2} \Delta r)$ converge to constants in the $\Delta r \ll r$ limit.

This is consistent in terms of the energy-momentum tensors,
\begin{gather}
T^{\mu}_{\;\;\nu} \simeq \left(
  \begin{array}{ccc}
    - \lambda & 0 & 0\\
    0 & +\lambda & 0 \\
    0 & 0 & +p \\
  \end{array}
\right) \delta(R-r) + \left(\mathrm{bulk\;terms}\right),
\end{gather}
and hence
\begin{gather}
\left. T^{\mu}_{\;\;\nu} \right|_{\mathrm{shell}} \simeq \left(
  \begin{array}{cc}
    - \lambda & 0\\
    0 & + p \\
  \end{array}
\right)
\end{gather}
on the shell, where
\begin{eqnarray}
\lambda &\equiv& \lambda_{\psi} + n^{2} \lambda_{a},\\
p &\equiv& -\lambda_{\psi} + n^{2} \lambda_{a},
\end{eqnarray}
and we define $\lambda_{\psi}$ and $\lambda_{a}$ are tension terms those came from the kinetic terms of $\psi$ and $a$, respectively. By solving the energy-conservation equation, we obtain
\begin{eqnarray}
\lambda_{\psi} = \mathrm{const}, \;\;\; \lambda_{a} = \frac{C}{r^{2}},
\end{eqnarray}
where $C$ is a constant. Therefore, it is consistent to match that $\psi_{\mathrm{f}}^{2}/\Delta r \simeq \lambda_{\psi}$ and $1/(e^{2} \Delta r) \simeq C$ in the thin-shell limit.

Then the junction equation what we have to solve is
\begin{eqnarray}
\epsilon_{-}\sqrt{\dot{r}^{2} + 1 + \frac{r^{2}}{\ell_{-}^{2}}} - \epsilon_{+}\sqrt{\dot{r}^{2} - M_{+} + \frac{r^{2}}{\ell_{+}^{2}}} = \left( \lambda_{\psi} + n^{2} \frac{C}{r^{2}} \right) r.
\end{eqnarray}
We emphasize again that this equation is consistent when $\psi_{\mathrm{f}}^{2}/\Delta r$ and $1/(e^{2} \Delta r)$ converge to constants in the $\Delta r \ll r$ limit. Physically, this requires $\psi_{\mathrm{f}} \ll 1$ (this can be assumed by tuning a shape of the potential $U$) and $e \gg 1$, i.e., the magnetic charge ($\propto 1/e$) should be sufficiently small.

\begin{figure}
\begin{center}
\includegraphics[scale=0.6]{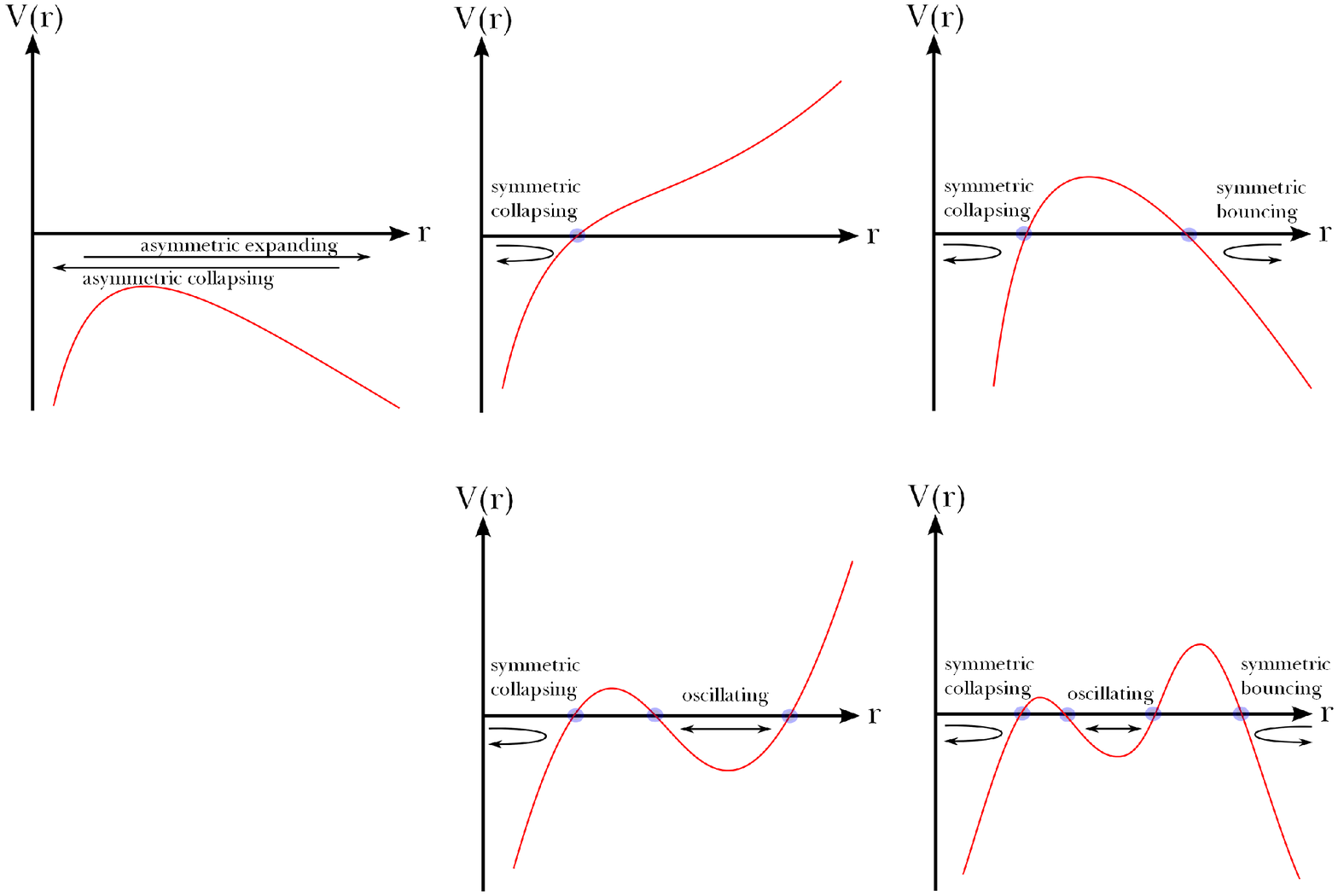}
\caption{\label{fig:pots}Possible classes of effective potentials $V(r)$.}
\includegraphics[scale=0.45]{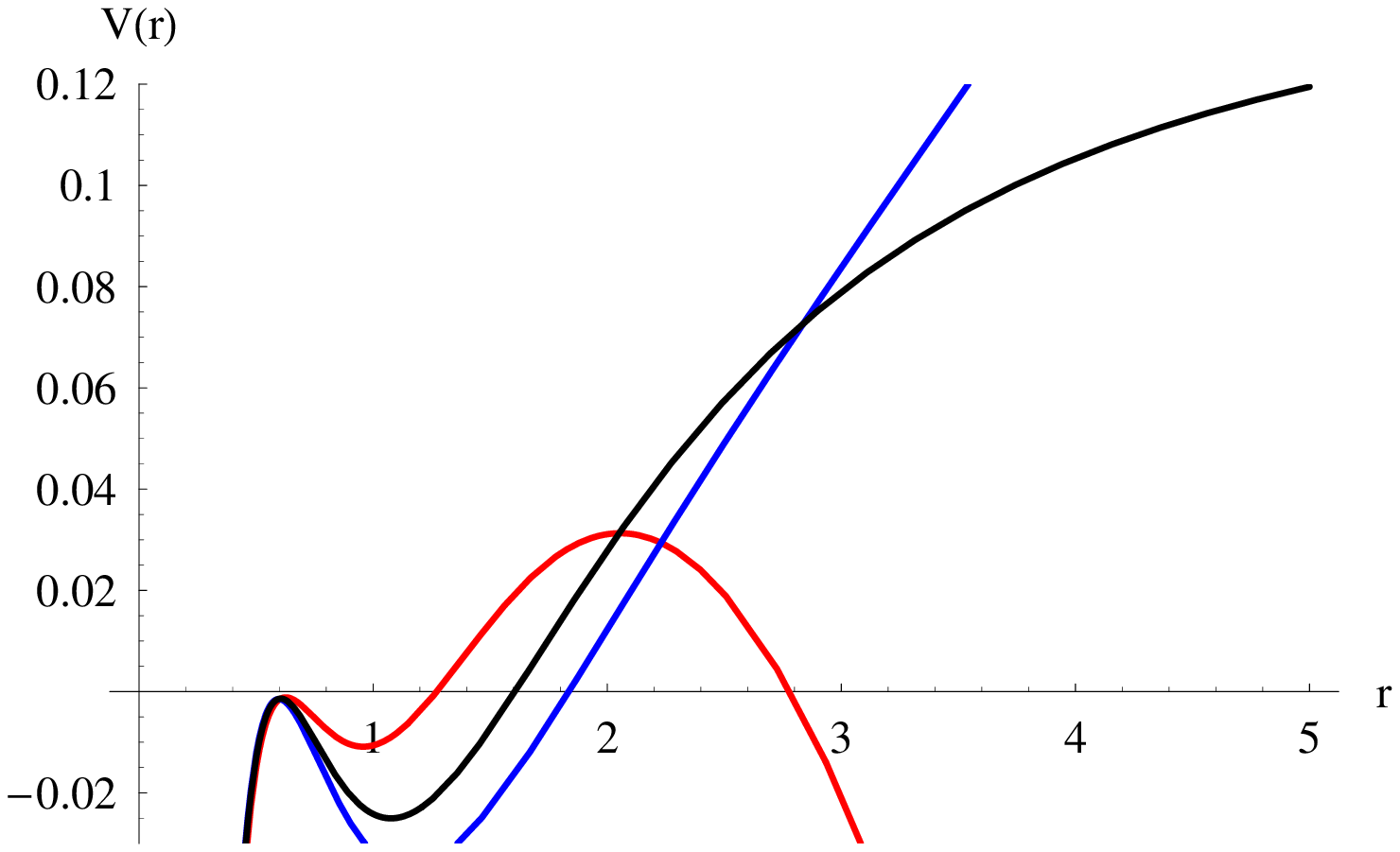}
\includegraphics[scale=0.45]{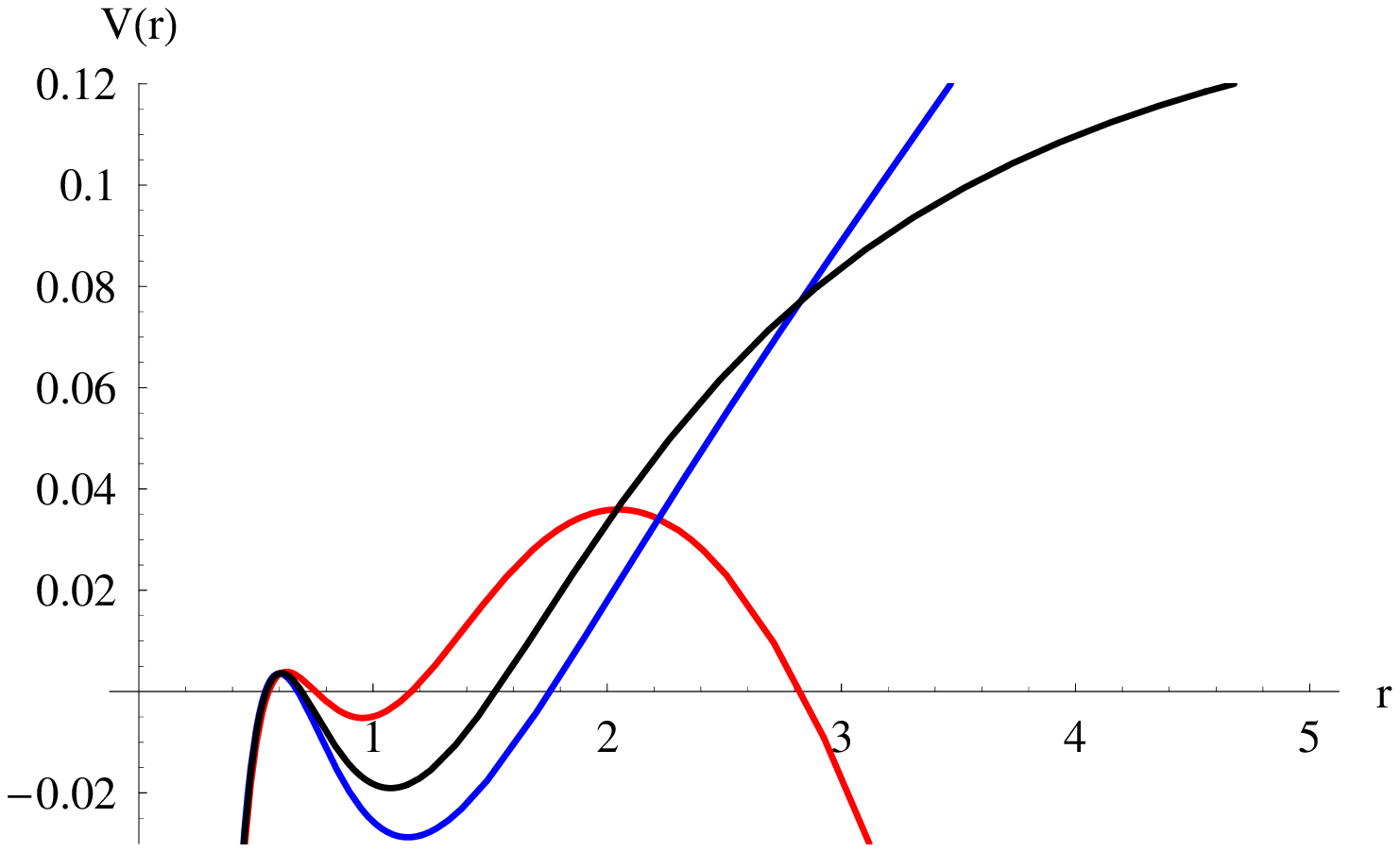}
\caption{\label{fig:zeros_onetwo}Examples of various zeros, where $\ell_{+}=10$, $\lambda_{\psi} = n^{2}C = 0.45$, $M_{+}=0.005$ (left, $r_{+}\simeq 0.7$) or $M_{+}=10^{-6}$ (right, $r_{+}\simeq 0.01$), and $\ell_{-} = 5$ (red, there exist bouncing solutions), $\ell_{-} = 2.5$ (blue, there is no bouncing solutions), and $\ell_{-} \simeq 2.86$ (black, critical limit).}
\end{center}
\end{figure}

\section{\label{sec:dyn}Dynamics of thin magnetic shells}

In this section, we classify possible classical trajectories of magnetic shells as well as their possible tunneling channels.

\subsection{Classification of effective potential}

First let us observe the possible solutions those satisfy $\dot{r}^{2} = 0$. From simple observations, it is trivial that $V(r) = 0$ is in fact a fourth order equation of $r^{2}$ (i.e., eighth order equation of $r$). Therefore, if we restrict $r > 0$, then there are at most four solutions.

In addition, let us see the behaviors of small $r$ and large $r$ limits.
\begin{itemize}
\item[--] In $r\rightarrow 0$ limit,
\begin{eqnarray}
V(r) \simeq - \frac{n^{2} C}{4 r^{2}} + \frac{1}{2}\left( 1 - M_{+} - \lambda_{\psi} n^{2} C \right) + \frac{1}{2} \left[ \left(\frac{1}{\ell_{-}^{2}} + \frac{1}{\ell_{+}^{2}} - \frac{\lambda_{\psi}^{2}}{2} \right) - \frac{\left( 1 + M_{+} \right)^{2}}{2C^{2} n^{2}} \right] r^{2} + \mathcal{O} \left( r^{4} \right).
\end{eqnarray}
Therefore, at once $C > 0$, then always $r$ goes to negative infinity.
\item[--] In $r \rightarrow \infty$ limit
\begin{eqnarray}
V(r) \simeq \left[\frac{1}{\ell_{+}^{2}} - \frac{1}{4 \lambda_{\psi}^{2}} \mathcal{A}^{2} \right] r^{2} + \left[ -M_{+} - \frac{\mathcal{A}}{2\lambda_{\psi}^{2}} \left(1 + M_{+} - \frac{n^{2}C}{\lambda_{\psi}} \mathcal{A} - 2 \lambda_{\psi} n^{2}C \right) \right] + \mathcal{O} \left( \frac{1}{r^{2}} \right),
\end{eqnarray}
where
\begin{eqnarray}
\mathcal{A} \equiv \frac{1}{\ell_{-}^{2}} - \frac{1}{\ell_{+}^{2}} - \lambda_{\psi}^{2}.
\end{eqnarray}
Therefore, the sign of $1/\ell_{+}^{2} - \mathcal{A}^{2}/4 \lambda_{\psi}^{2}$ determines the behavior of $r$ in the large $r$ limit, and hence if it is positive, zero, or negative, then $V(r)$ diverges to positive infinity, converges to a constant, or diverges to negative infinity, respectively.
\end{itemize}

\begin{figure}
\begin{center}
\includegraphics[scale=0.6]{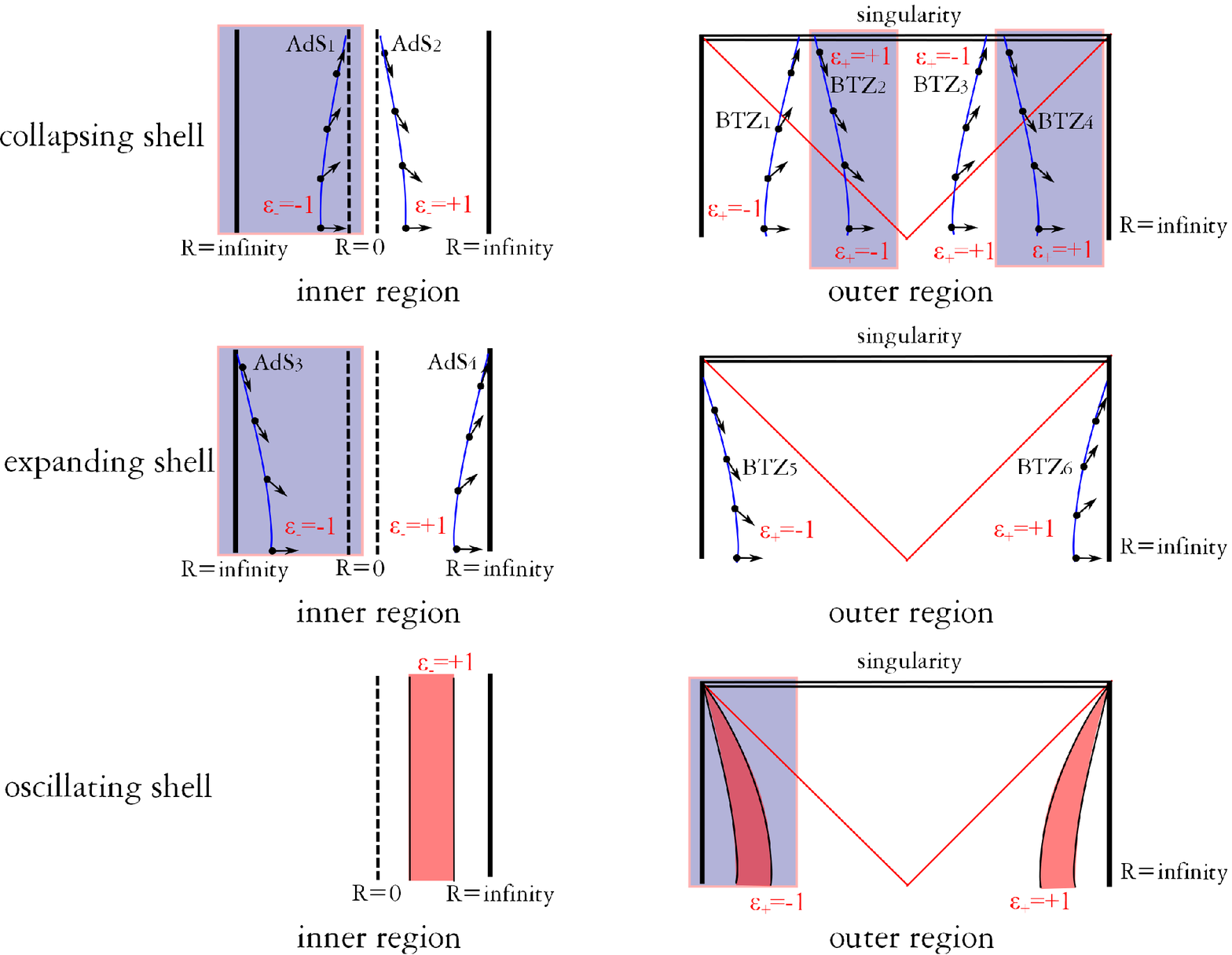}
\caption{\label{fig:signs} $\epsilon_{\pm}$ and their trajectories for symmetric and oscillating solutions. The gray colored boxes are not allowed. Small black arrows denote outward normal directions.}
\includegraphics[scale=0.6]{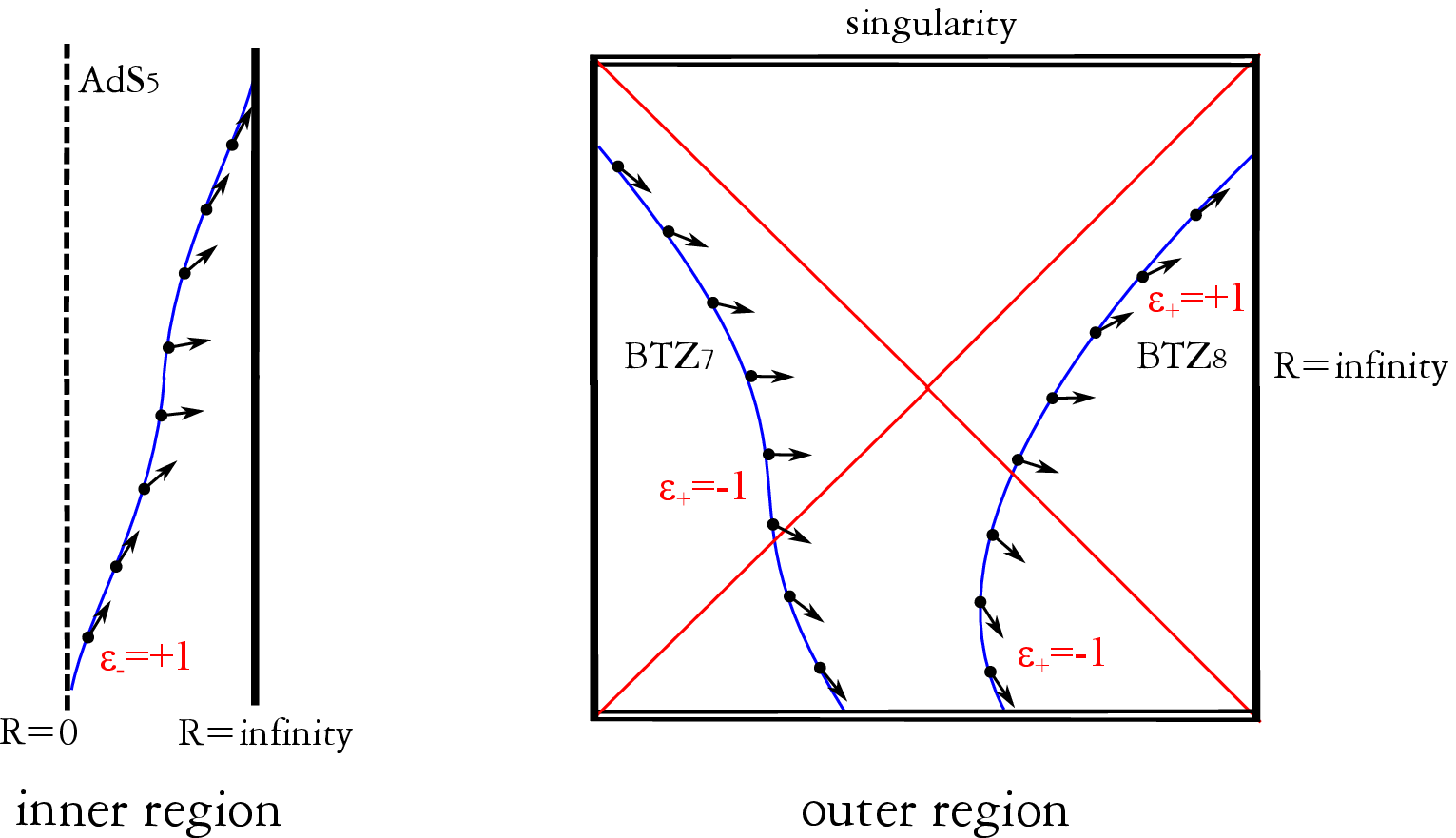}
\caption{\label{fig:signs2} $\epsilon_{\pm}$ and their trajectories for asymmetric solutions.}
\end{center}
\end{figure}

Therefore, there are possible cases depending on the number of zeros of $V(r) = 0$ (FIG.~\ref{fig:pots}; for numerical examples, FIG.~\ref{fig:zeros_onetwo}):
\begin{itemize}
\item[1.] \textit{No zero}. Then there are \textit{asymmetric} collapsing and expanding solutions.
\item[2.] \textit{One zero}. Then there is only a \textit{symmetric} collapsing solution.
\item[3.] \textit{Two zeros}. Then there are \textit{symmetric} collapsing and bouncing solutions.
\item[4.] \textit{Three zeros}. Then there is a \textit{symmetric} collapsing solution, as well as an \textit{oscillating} solution.
\item[5.] \textit{Four zeros}. Then there are \textit{symmetric} collapsing and bouncing solutions, as well as an \textit{oscillating} solution.
\end{itemize}
Asymmetric and symmetric solutions are already well-known even in four dimensional cases. However, the oscillating solution is quite a new feature in our model (although an oscillating solution was known from a space-like shell \cite{Balbinot:1990zz} or a rotating shell \cite{Mann:2006yu}).

\subsection{Causal structures and tunneling channels}

The extrinsic curvatures $\beta_{\pm}$ determine the signs of $\epsilon_{\pm}$, where
\begin{eqnarray}
\beta_{\pm} &\equiv& \frac{f_{-} - f_{+} \mp \lambda^{2} r^{2}}{2\lambda r} = \epsilon_{\pm} \sqrt{\dot{r}^{2} + f_{\pm}}\\
&\propto&
\left\{ \begin{array}{ll}
\mp \frac{n^{2}}{2} C & \;\;\;\;\; r\rightarrow 0,\\
\frac{1}{\ell_{-}^{2}}- \frac{1}{\ell_{+}^{2}} \mp \lambda_{\psi}^{2} & \;\;\;\;\; r\rightarrow \infty.
\end{array} \right.
\end{eqnarray}
Therefore, for the collapsing shell, $\epsilon_{+} = -1$ and $\epsilon_{-} = +1$ in the $r\rightarrow 0$ limit. In addition, for the bouncing shell, it is up to the signs of $\mathcal{A}$ and $\mathcal{A} + 2\lambda_{\psi}^{2}$. Note that for the inner region, $\beta_{-} < 0$ is beyond the physical region and hence it is more reasonable to choose $\mathcal{A} + 2\lambda_{\psi}^{2} > 0$.

From these information, let us classify classical solutions (FIG.~\ref{fig:signs} and FIG.~\ref{fig:signs2}):
\begin{itemize}
\item[--] For a symmetric collapsing solution, there are two possibilities, either $\epsilon_{+}$ is always negative and hence $\mathrm{AdS}_{2} - \mathrm{BTZ}_{1}$ or $\epsilon_{+}$ changes the sign and hence $\mathrm{AdS}_{2} - \mathrm{BTZ}_{3}$.
\item[--] For a symmetric bouncing solution, there are two possibilities, either $\epsilon_{+}$ is negative and hence $\mathrm{AdS}_{4} - \mathrm{BTZ}_{5}$ or $\epsilon_{+}$ is positive and hence $\mathrm{AdS}_{4} - \mathrm{BTZ}_{6}$.
\item[--] For an oscillating solution, the shell should not cross over the event horizon. Hence, $\beta_{+}$ cannot change the sign. What we are interested in is the case $\beta_{\pm} >0$ for all cases.
\item[--] For an asymmetric solution, if $\beta_{+}$ does not change the sign, then it becomes $\mathrm{AdS}_{5} - \mathrm{BTZ}_{7}$, while if it changes, then it becomes $\mathrm{AdS}_{5} - \mathrm{BTZ}_{8}$. In addition, their opposite diagrams (asymmetric collapsing solutions) are also allowed.
\end{itemize}

\begin{figure}
\begin{center}
\includegraphics[scale=0.55]{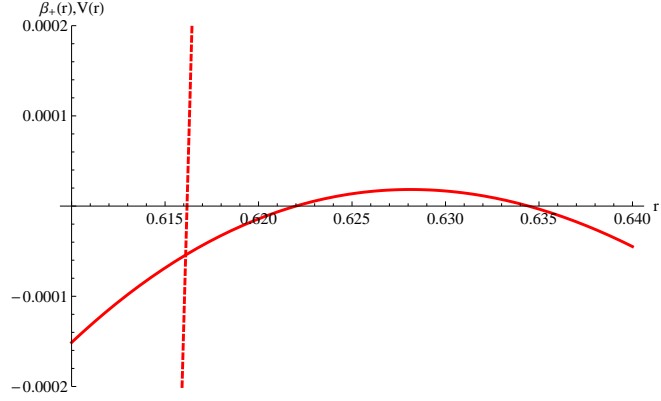}
\caption{\label{fig:near_extreme} $\beta_{+}$ (red dashed) and $V(r)$ (red) for $\ell_{+}=10$, $\lambda_{\psi} = n^{2}C = 0.45$, $M_{+}=0.00385$, and $\ell_{-} = 5$. There exists a symmetric collapsing solution that has a positive $\beta_{+}$ around the symmetric point.}
\end{center}
\end{figure}
\begin{figure}
\begin{center}
\includegraphics[scale=0.45]{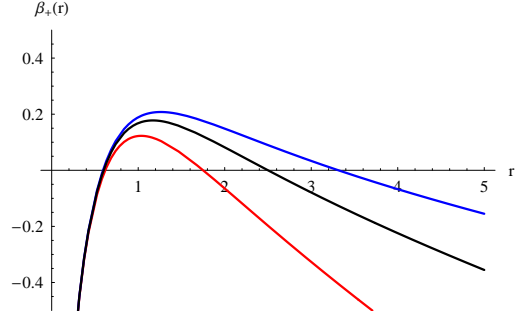}
\caption{\label{fig:betap} $\beta_{+}$ for $\ell_{+}=10$, $\lambda_{\psi} = n^{2}C = 0.45$, $M_{+}=10^{-6}$, and $\ell_{-} = 5$ (red), $\ell_{-} = 2.5$ (blue), and $\ell_{-} \simeq 2.86$ (black). Note that $\beta_{-}$ is positive for all region.}
\end{center}
\end{figure}

Let us tune parameters so that there exists $r_{0}$ with $V(r_{0}) = V'(r_{0}) = 0$. Note that $V = f_{+} - \beta_{+}^{2}$ and $V' = f'_{+} - 2 \beta_{+} \beta'_{+}$. Near $r_{0}$, $r$ should be larger than the size of the event horizon and hence it should be $f_{+} > 0$. Therefore, $\beta_{+} > 0$ is required. Like this, if we tune parameters around this extreme limit, then we can find a symmetric collapsing solution that has a region where $\beta_{+} > 0$ around the turning point ($V(r)=0$), e.g., FIG.~\ref{fig:near_extreme}. On the other hand, it is also possible that $\beta_{+}$ is always negative for symmetric collapsing solutions, while $\beta_{+}$ is always positive for oscillating solutions (e.g., FIG.~\ref{fig:betap}). For convenience, we will only consider the case that $\beta_{+} < 0$ for all symmetric bouncing solutions.

\begin{figure}
\begin{center}
\includegraphics[scale=0.6]{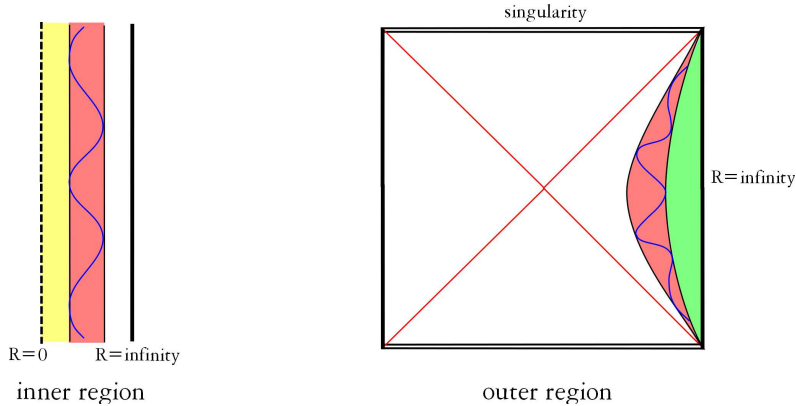}
\caption{\label{fig:tunneling_shell}The causal structure of an oscillatory magnetic shell (left is inside and right is outside).}
\end{center}
\end{figure}

\begin{figure}
\begin{center}
\includegraphics[scale=0.6]{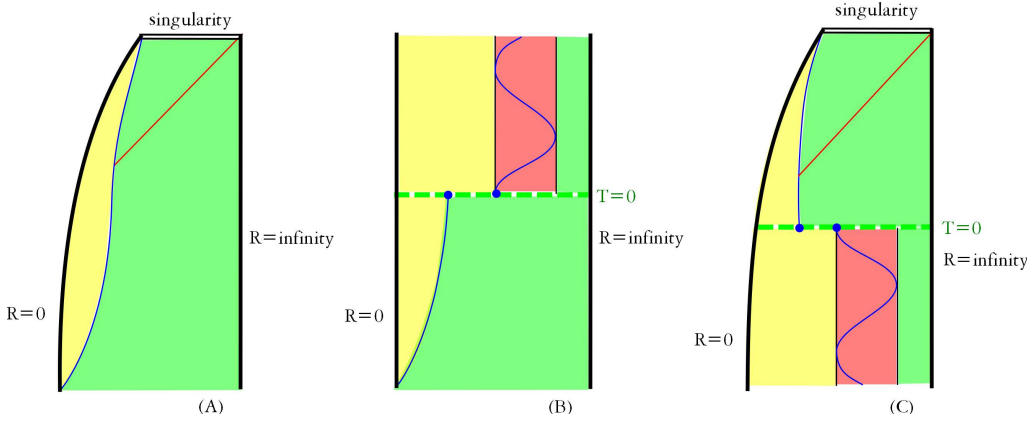}
\caption{\label{fig:tunneling3} Possible tunneling channels when the shell does not cross the event horizon. (A) A regular process generates a collapsing shell. (B) A collapsing shell can tunnel and form an oscillating shell. (C) An oscillating shell tunnels to a collapsing shell and forms a black hole.}
\end{center}
\end{figure}

Then by choosing parameters, the following tunneling channels are possible via Farhi-Guth-Guven tunneling \cite{Farhi:1989yr}. FIG.~\ref{fig:tunneling_shell} is the causal structure of an oscillating shell. This shell can be built by a regular initial condition, or it can be tunneled from a regular initial collapsing shell ((B) in FIG.~\ref{fig:tunneling3}), while it was originally just a collapsing shell that forms a black hole ((A) in FIG.~\ref{fig:tunneling3}, analogous with \cite{Gregory:2013hja}). For (A) and (B) in FIG.~\ref{fig:tunneling3}, we did not specified the initial singularity, since it does not suffer from a singularity theorem and we may think that a physical condition can induce such a gravitational collapse, like a star interior. Note that (A) is possible only if the parameters allow $\beta_{+} > 0$ for a certain region of the symmetric collapsing solution part.

\begin{figure}
\begin{center}
\includegraphics[scale=0.6]{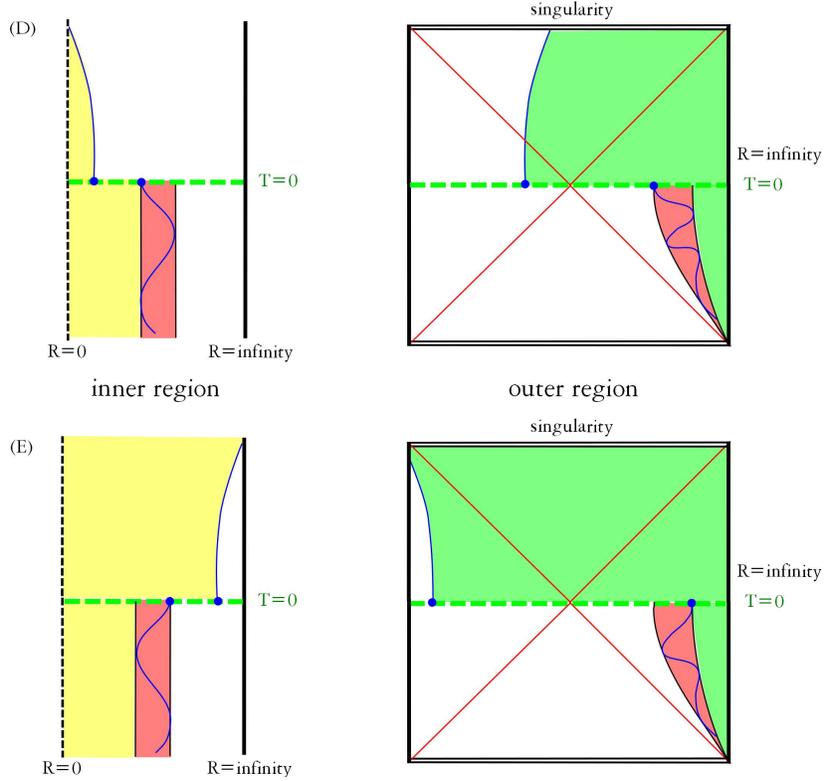}
\caption{\label{fig:tunneling4} Possible tunneling channels when the shell crosses the event horizon. (D) An oscillating shell tunnels to a collapsing shell and forms a black hole. (E) An oscillating shell tunnels to an expanding shell.}
\end{center}
\end{figure}

After an oscillating shell is formed, it can tunnel to either a collapsing or an expanding shell. If it allows a $\beta_{+} > 0$ region for a symmetric collapsing shell, then (C) in FIG.~\ref{fig:tunneling3} is the causal structure after the tunneling; if $\beta_{+} < 0$ for all region, then it should tunnel as (D) in FIG.~\ref{fig:tunneling4}. If the oscillating shell tunnels to an expanding shell and $\beta_{+} < 0$ for such an expanding region, then (E) in FIG.~\ref{fig:tunneling4} is the corresponding causal structure. For this case, the shell will touch the future infinity and it will form a Cauchy horizon in the end.


\subsection{Comments on information loss problem}

Let us imagine a situation that, from a unitary process, an oscillatory shell was created. If it tunnels to smaller or larger shell, then it should form a BTZ black hole. As we know, a BTZ black hole does not completely evaporate and hence the black hole is eternal. In terms of correlations between the inside the black hole and the asymptotic observer, it will eventually decay to zero as time goes on. Therefore, for a fixed BTZ geometry, information disappears.

On the other hand, at once there is a probability to form a magnetic shell, there is some hope to recover information. For a given asymptotic anti de Sitter boundary, in fact, there are infinitely many numbers of tunneling channels, and the unitary observer should add all geometries to obtain the correct correlations. The effective geometry of the observer would look like $\langle g_{\mu\nu} \rangle$, where this can be approximated by summing over on-shell histories:
\begin{eqnarray}
\langle g_{\mu\nu} \rangle \simeq p_{\mathrm{c}} g_{\mu\nu}^{(\mathrm{c})} + \left( p_{\frac{1}{2}} g_{\mu\nu}^{(\frac{1}{2})} + p_{1} g_{\mu\nu}^{(1)} + p_{\frac{3}{2}} g_{\mu\nu}^{(\frac{3}{2})} + ... \right),
\end{eqnarray}
where $g_{\mu\nu}^{(\mathrm{c})}$ is the oscillatory magnetic shell (left and right of FIG.~\ref{fig:tunneling_shell}), $g_{\mu\nu}^{(\frac{1}{2})}$ is the geometry that tunnels from the maximal radius of the oscillating shell to the expanding shell ((E) of FIG.~\ref{fig:tunneling4}) after a half period oscillation, $g_{\mu\nu}^{(1)}$ is the geometry that tunnels from the minimum radius of the oscillating shell to the collapsing shell ((C) of FIG.~\ref{fig:tunneling3} or (D) of FIG.~\ref{fig:tunneling4}) after one period oscillation, etc. Note that $p_{i}$s are probabilities of corresponding histories, i.e., $p_{\mathrm{c}}$ is the probability to form an oscillating shell, $p_{\frac{n}{2}}$ is the probability that tunnels from an oscillating shell to an expanding shell after $n/2$ oscillations, and $p_{n}$ is the probability that tunnels from an oscillating shell to a collapsing shell after $n$ oscillations, for each $n$.

In terms of correlation functions, the correlation through the geometry $g_{\mu\nu}^{(\mathrm{c})}$ will not decay in the end via the trivial geometry. Moreover, as $n > 0$ increases, even though geometries $g_{\mu\nu}^{(n)}$ or $g_{\mu\nu}^{(\frac{n}{2})}$ will eventually form a black hole, they will sustain a trivial geometry for a finite but sufficiently long time so that correlations could be recovered to the boundary. Therefore, through such histories, correlations will be recovered and hence even though the probability is low, they will be dominated after a sufficiently long time. This is the way to recover the correlations and unitarity of the boundary observer.

Note that each geometry will be classicalized independently. Therefore, if one observer freely falls into the black hole and follows the most probable history, then the observer will eventually lose information. On the other hand, the unitary observer who can sum all histories will eventually recover information. This is one realization of the idea so-called \textit{effective loss of information}. One interesting point is that, regarding our new example, there will be an infinite number of histories that contribute to unitarity. This is also worthwhile to note that the unitary observer should gather all geometries and hence the usual Einstein equation cannot hold for $\langle g_{\mu\nu} \rangle$. This is the reason why the unitarity and general relativity seemed inconsistent according to \cite{Almheiri:2012rt}. Therefore, in some sense, $\langle g_{\mu\nu} \rangle$ represents an effective realization of the firewall, and it is fair to say that we do not need the physical firewall as a physically singular surface around the horizon scale.

\section{\label{sec:con}Conclusion}

In this paper, we investigated dynamics of a thin-magnetic shell in three dimensional anti de Sitter space. We were interested in the case that the system allows an oscillating shell. It can tunnel to either a collapsing shell or an expanding shell. For both cases, they will form a singularity and an event horizon, and hence information will be lost. However, still there exists a possibility to maintain the trivial geometry via a meta-stable oscillating shell. Therefore, even though there are solutions which result information loss, after the path integral, the entire wave function will conserve information.

This is a good realization of the effective loss of information. Our example has some interesting properties:
\begin{itemize}
\item[--] We obtained a time-like oscillating shell. This is due to the magnetic field, while the electric field \cite{Alberghi:1999kd} or any other typical gravitational effects (mass, vacuum energy, etc.) were difficult to allow this oscillatory shell.
\item[--] The trivial geometry can tunnel to infinitely divergent histories. Therefore, the unitary observer should sum over infinitely many histories.
\end{itemize}
Our example may imply that not only gravity, but also some special properties of gauge fields can contribute to the information loss problem. In addition, it opens a possibility that contributions from gauge fields as well as their non-perturbative effects (vortex, magnetic monopole, etc.) will do an important role to the information loss problem; and these will make a complication to the entire path integral. 

Can our result be generalized, for example, to four dimensional spacetime? Can it be applied to Minkowski or de Sitter background? Can there be an analogous model that can be embedded by the standard model, GUT, or any other particle physics motivated models? We open these possibilities for future projects.

\newpage

\section*{Acknowledgment}
DY was supported by Leung Center for Cosmology and Particle Astrophysics (LeCosPA) of National Taiwan University (103R4000). BHL was supported by the National Research Foundation of Korea (NRF) grant funded by the Korea government (MSIP, 2014R1A2A1A01002306). WL was supported by Basic Science Research Program through the National Research Foundation of Korea (NRF) funded by the Ministry of Education (2012R1A1A2043908). WL would like to thank Hongsu Kim for helpful discussions and comments.

\end{document}